\documentclass[preprint,12pt,3p]{elsarticle}

\usepackage{amssymb}
\usepackage{graphicx}
\usepackage{amsmath}
\usepackage{caption}

\journal{Sample Journal}

\newtheorem{theorem}{Theorem}

\newtheorem{example}{Example}

\newtheorem{problem}{Problem}

\newenvironment{proof}{ {\bf Proof:}} {$\Box$}

\begin{document}

\begin{frontmatter}

\title{A tight lower bound for the hardness of clutters}

\author{Vahan Mkrtchyan}
\ead{vahanmkrtchyan2002@ysu.am}

\author{Hovhannes Sargsyan\corref{cor1}}
\ead{hsargsian@gmail.com}
\cortext[cor1]{Corresponding author}
\address{Department of Informatics and Applied Mathematics, Yerevan State University, Yerevan, 0025, Armenia}







\begin{abstract}
A {\it clutter} (or {\it antichain} or {\it Sperner family}) $L$ is a pair $(V,E)$, where $V$ is a finite set and $E$ is a family of subsets of $V$ none of which is a subset of another. Normally, the elements of $V$ are called {\it vertices} of $L$, and the elements of $E$ are called {\it edges} of $L$. A subset $s_e$ of an edge $e$ of a clutter is {\it recognizing} for $e$, if $s_e$ is not a subset of another edge. The {\it hardness} of an edge $e$ of a clutter is the ratio of the size of $e\textrm{'s}$ smallest recognizing subset to the size of $e$. The hardness of a clutter is the maximum hardness of its edges. In this short note we prove a lower bound for the hardness of an arbitrary clutter. Our bound is asymptotically best-possible in a sense that there is an infinite sequence of clutters attaining our bound.
\end{abstract}


\end{frontmatter}



\section{Introduction}
\label{sec:intro}

A {\it clutter} (or {\it antichain} or {\it Sperner family}) $L$ is a pair $(V,E)$, where
$V$ is a finite set and $E$ is a family of subsets of $V$ none of
which is a subset of another. Following \cite{Cornuejols}, the
elements of $V$ will be called {\it vertices} of $L$, and the elements
of $E$ are called {\it edges} of $L$.

For a clutter $L=(V,E)$, a subset $e_0\subseteq e$ of an edge $e$ is a
{\it recognizing} subset for $e$, if $e_0\subseteq e'$ for some $e'\in E$, then $e'=e$. Let $s_e$ be a smallest recognizing subset of $e\in E$. Define $c(e)=|s_{e}|/|e|$, and let
\begin{equation*}
c(L)=\max_{e \in E}\textrm{ }c(e).
\end{equation*}

$c(L)$ is called the {\it hardness} of $L$. Note that $0\leq c(L)\leq 1$ for any clutter $L=(V,E)$. Moreover, if $|E|\leq 1$, then clearly $c(L)=0$. Thus, we will consider clutters $L$ with at least two edges. In this case any edge contains no more than $|V|-1$ vertices and any recognizing subset of an edge $e\in E$ contains at least one vertex. Hence,
\[\frac{1}{|V|-1}\leq c(L)\leq 1.\]
The lower bound is asymptotically tight as the following example demonstrates.

\begin{example}\label{exmp:1/|V|-1} For any positive integer $n$ ($n\geq 3$) let $V_n=\{1,...,n\}$, $E_n=\{e_1, e_2\}$, where $e_1=\{1,3,...,n\}$ and $e_2=\{2,3,...,n\}$. Then $L_n=(V_n,E_n)$ is a clutter. The set $\{1\}$ is recognizing for $e_1$, and $c(e_1)=\frac{1}{n-1}$. Similarly, the set $\{2\}$ is recognizing for $e_2$, and $c(e_2)=\frac{1}{n-1}$. Hence
\[c(L_n)=\max\{c(e_1), c(e_2)\}=\frac{1}{n-1}=\frac{1}{|V_n|-1}.\]
\end{example}
The main reason why the clutter $L_n$ has such a low hardness, is that the elements $3,...,n$ are present in every edge. This means that they cannot be present in any smallest recognizing subset of an edge. Therefore they do not contribute to the numerator of the hardness of an edge, however, they do contribute to its denominator. This situation prompts the following
\begin{problem}\label{MainProblem} Find a best-possible function $f$ such that any clutter $L=(V,E)$ satisfying the condition
\begin{enumerate}
	\item [(C1)] no vertex of $L$ is present in all edges of $L$,
\end{enumerate}has hardness $c(L)\geq f(|V|)$.
\end{problem}Our considerations above imply that $f(|V|)\geq \frac{1}{|V|-1}$. Note that any clutter satisfying (C1) has at least two edges, moreover, without loss of generality, we can assume that the clutters in the formulation of the Problem \ref{MainProblem} satisfy
\begin{enumerate}
	\item [(C2)] each vertex of $L$ is present in at least one edge of $L$.
\end{enumerate}This follows from the observation that isolated vertices (vertices, that do not belong to an edge) can be removed from the clutter without affecting its hardness.

In \cite{OpusculaComplexity}, the Problem \ref{MainProblem} is addressed for two classes of clutters that arise from graphs. Let us note that the graphs considered in this paper (and in \cite{OpusculaComplexity}) are finite, undirected and do not contain multiple edges or loops. Formally, such a graph $G$ can be considered as a clutter $(V,E)$, in which $E$ is any subset of the set of pairs of elements from $V$.

For a graph $G$, let $V(G)$ and $E(G)$ be the sets of vertices and edges of $G$, respectively. There is an important comment that should be made here concerning the terminology. An edge of a clutter is a subset of the set of vertices, and therefore it can contain more than two vertices, however, an edge of a graph contains exactly two vertices. 

A set $V'\subseteq V(G)$ is said to be {\it independent}, if $V'$ contains no adjacent vertices. Similarly, $E'\subseteq E(G)$ is independent, if $E'$ contains no adjacent edges. An independent set of vertices (edges) is called {\it maximal}, if it does not lie in a larger independent set. An independent set of edges is also called {\it matching}.

Independent sets give rise to clutters. If for a graph $G=(V,E)$ we denote the set of all maximal independent sets of vertices of $G$ by $U_{G}$, then $(V,U_{G})$ is a
clutter. In the paper we use $\mathcal{U}_G$ to denote the clutter $(V,U_G)$. 

Another clutter that a graph $G=(V,E)$ gives rise is $(E,M_G)$, where $M_G$ denotes the set of all maximal matchings of $G$. This clutter will be denoted by $\mathcal{M}_G$.

In \cite{OpusculaComplexity}, $c(\mathcal{U}_G)$ and $c(\mathcal{M}_G)$ are investigated. In particular, it is shown that 
\begin{equation*}
c(\mathcal{U}_G)\geq \frac{1}{1+|V(G)|-2\sqrt{|V(G)|-1}}
\end{equation*}
provided that $G$ is a connected graph different from $K_1,K_{2,2},K_{3,3},K_{4,4}$. Here $K_n$ denotes the complete graph on $n$ vertices. Moreover, $K_{m,n}$ is the complete bipartite graph on $m+n$ vertices, such that one side has $m$ vertices and the other side $n$ vertices. In \cite{OpusculaComplexity}, it is shown that this bound is attained by infinitely many graphs. We observe that $\mathcal{U}_G$ satisfies conditions (C1) and (C2) when $G$ is a connected graph containing at least $2$ vertices. This follows from an observation that a vertex forms an independent set in $G$, therefore it can be extended to a member of $\mathcal{U}_G$. Since $G$ is connected, no vertex of $G$ belongs to all members of $\mathcal{U}_G$. Note that this implies that in the search of the function $f$ for Problem \ref{MainProblem}, one should restrict herself/himself exclusively to those functions that satisfy the following inequality:
\begin{equation*}
\frac{1}{|V|-1}\leq f(|V|)\leq \frac{1}{1+|V|-2\sqrt{|V|-1}}.
\end{equation*} 

The aim of this short note is to give a best-possible answer to Problem \ref{MainProblem}. Our Theorem \ref{thm:Main} presents a lower bound for the hardness of any clutter that satisfies conditions (C1) and (C2). Moreover, the bound is asymptotically tight in a sense that there is an infinite sequence of clutters attaining the bound.

Finally, let us note that the hardness of a clutter, that is introduced in \cite{OpusculaComplexity}, is new (see \cite{Claus,Jukna} where the authors introduce two different types of hardness for graphs). Terms and concepts that we do not define can be found in \cite{Cornuejols,West}.

\section{The main result}
\label{sec:Main}

In this section we present the main result of the paper. As Example \ref{exmp:1/|V|-1} demonstrates, $\frac{1}{|V|-1}$ is the best bound that we can obtain for general clutters with at least $2$ edges. However, if we consider the clutters that satisfy (C1) and (C2), the lower bound for the hardness of such clutters can be improved.

\begin{theorem}\label{thm:Main} Let $L=(V,E)$ be a clutter satisfying the conditions (C1) and (C2). Then \[c(L)\geq \frac{1}{|V|-2\sqrt{|V|}+2}.\]
\end{theorem}
\begin{proof} Assume that $|V|=n$ and $|E|=m$. If there is an edge $e\in E$ with $|s_e|\geq 2$, then taking into account that $|e|\leq n-1$, we have
\[c(L)\geq \frac{2}{n-1}\geq \frac{1}{n-2\sqrt{n}+2}.\]
Thus, without loss of generality, we can assume that for each edge  $e\in E$ $|s_e|=1$.

Define a graph $G=(V,X)$ on vertices of $L$ as follows: for two vertices $u, v\in V$, we have $(u,v)\in X$, if and only if for each $e\in E$ $e$ does not contain both $u$ and $v$ ($\{u,v\}\nsubseteq e$).

If $E=\{e_1,..., e_m\}$, then by definition of $G$, we have that the unique vertices that belong to $s_{e_1},..., s_{e_m}$ form an $m$-clique $Q$ in $G$. Let us show that each vertex $w$ lying outside $Q$ is adjacent to a vertex of $Q$. If $w$ is not adjacent to any of vertices of $Q$, then by definition of $G$, it belongs to all edges $e_1,..., e_m$ of $L$ violating condition (C1).

Since the number of vertices lying outside $Q$ is $n-m$, we have that at least one vertex $z$ of $Q$ has degree at least $d(z)\geq m-1+\frac{n-m}{m}=m+\frac{n}{m}-2$ in $G$. Let $e_z$ be the unique edge of $L$ containing $z$. We have
\[|e_z|\leq n-d(z)\leq n-(m+\frac{n}{m})+2,\]
therefore
\[c(L)\geq c(e_z)\geq \frac{1}{n-(m+\frac{n}{m})+2}.\]

Observe that since any edge of $L$ contains a unique vertex, we have $m\leq n$. Consider the function $f(x)=x+\frac{n}{x}$ on the interval $[1,n]$. Using the elementary calculus, it can be shown that
\[\min_{x\in [1,n]}f(x)=2\sqrt{n} \text{ and }\max_{x\in [1,n]}f(x)=n+1.\]

Hence
\[m+\frac{n}{m}\geq 2\sqrt{n},\]
or
\[n-(m+\frac{n}{m})+2\leq n-2\sqrt{n}+2.\]
Since $m+\frac{n}{m}\leq n+1$, we have
\[n-(m+\frac{n}{m})+2>0,\]
hence
\[c(L)\geq \frac{1}{n-(m+\frac{n}{m})+2} \geq \frac{1}{n-2\sqrt{n}+2}.\]

The proof of the theorem is complete.
\end{proof}

\smallskip

Let us note that there is an infinite sequence of clutters achieving the bound of the Theorem \ref{thm:Main}. This example is considered in \cite{OpusculaComplexity}. Let $k$ be any positive integer with $k\geq 2$. Take $n=k^2$, and let $U_0$ be a set with $n-k=k(k-1)$ elements. Consider an $n$-vertex graph $G$ obtained from a $k$-clique $Q$, by joining every vertex of $Q$ to $k-1$ elements of $U_0$ (each element of $U_0$ is joined to exactly one vertex of $Q$). Note that $U_0\in U_G$. Let $L$ be the clutter  that is obtained from $\mathcal{U}_G$ by removing the edge $U_0$. Observe that all edges of $L$ contain exactly $1+(k-1)^2$ vertices, moreover, a set comprised of a vertex of $Q$ is a smallest recognizing subsets for an edge of $L$. Thus
\[c(L)=\frac{1}{1+(k-1)^2}=\frac{1}{n-2\sqrt{n}+2}.\]
We observe that for each $k\geq 3$ there is no connected graph $G$, such that $\mathcal{U}_G$ coincides with $L$ constructed above. This follows from an observation that for clutters $\mathcal{U}_G$, where $|V(G)|=n$, $G$ is connected and $G$ is different from $K_1,K_{2,2},K_{3,3},K_{4,4}$, we have the bound $c(\mathcal{U}_G)\geq \frac{1}{1+n-2\sqrt{n-1}}$ \cite{OpusculaComplexity} which exceeds $\frac{1}{n-2\sqrt{n}+2}$. Finally, one can easily see that when $k=2$, $L=\mathcal{U}_{K_{2,2}}$.

%



\bibliographystyle{elsarticle-num}


\end{document}